\documentclass[11pt,twoside]{article}

\usepackage{baaa2013}

\usepackage{graphicx}
\usepackage{subfigure}
\usepackage{amssymb}
\usepackage[spanish,activeacute,english]{babel}
\usepackage[latin1]{inputenc}
\usepackage{verbatim}
\usepackage{amsmath}
\usepackage{amsfonts}
\usepackage{amssymb}
\usepackage[colorlinks=true,dvips]{hyperref}
\usepackage{natbib}

\begin{document}
%%%%%%%%%%%%%%%%%%%%%%%%%%%%%%%%%%%%%%%%%%%%%%%%%%%%%%%%%%%%%%%%%%%%%%%%%%
%%%% SELECCIONE EL IDIOMA EN QUE SE ESCRIBE EL ARTÍCULO:              %%%%
\myselectspanish
\myselectenglish
%%%%%%%%%%%%%%%%%%%%%%%%%%%%%%%%%%%%%%%%%%%%%%%%%%%%%%%%%%%%%%%%%%%%%%%%%%
%%\vskip 1.0cm
%%\markboth{Rovero et al.}%
%%{Ejemplos e instrucciones para el BAAA}

\pagestyle{myheadings}
%%%% DESCOMENTE LA LINEA QUE DESCRIBE EL CARACTER DE SU TRABAJO       %%%%
\vspace*{0.5cm}

\Categ{1}
\Tema{7}

\vskip 0.3cm
\title{Search for extended gamma ray emission in Markarian 421 using VERITAS observations}

%\title{ Template paper for publication in the Bulletin of the 
%Argentinian Astronomical Association with instructions for the use of 
%\LaTeX{}}

\author{M. Fernandez Alonso$^{1}$ for the VERITAS Collaboration}

\affil{%
  (1) Instituto de Astronom\'ia y F\'isica del Espacio (CONICET-UBA)\\ 
  }

\begin{abstract} 
Very high energy (VHE: $>$100 GeV) gamma rays coming from AGN can pair-produce on the intergalactic background light generating an electromagnetic cascade. If the Intergalactic Magnetic Field (IGMF) is sufficiently strong, this cascade may result in an extended isotropic emission of photons around the source, or {\it halo}. Using VERITAS observations of the blazar Markarian 421, we search for extended emission by comparing the source angular distribution ($\theta^2$) from a quiescent period with one coming from a flare period, which can be considered as ``halo-free". $\chi^2$ test showed no significant statistical differences between the samples, suggesting that the effect is either non-existent or too weak to be detected. We calculated upper limits for the extended flux considering different angle ranges, the most stringent being $<$8\% of the Crab Nebulae flux (C.U), in the range $0^{o}\leq\theta\leq\ 0.1^{o}$.
\end{abstract}

\begin{resumen}
Los rayos gamma de muy altas energias (VHE: $>$100 GeV) provenientes de AGN pueden interactuar con fotones de los campos de radiaci\'on intergal\'acticos, generando una cascada electromagn\'etica. Si el Campo Magn\'etico Intergal\'actico es suficientemente intenso, esta cascada puede resultar en una emisi\'on isotr\'opica de fotones alrededor de la fuente, o {\it halo}. Utilizando observaciones del blazar Markarian 421 realizadas con VERITAS, se busca detectar emisi\'on extendida comparando la distribuci\'on angular de la fuente ($\theta^2$) proveniente de un per\'iodo de baja actividad con otra proveniente de un per\'iodo de flare, en la que la componente del halo puede considerarse despreciable. No se encontraron diferencias estad\'isticamente significativas entre las distribuciones de acuerdo al test de $\chi^2$, sugiriendo que el efecto es muy d\'ebil, o no est\'a presente en absoluto. Se calcularon cotas para el flujo de emisi\'on extendida considerando diferentes rangos de \'angulos, siendo la menor de las cotas de $<$8\% del flujo de la Nebulosa del Cangrejo (C.U) en el rango $0^{o}\leq\theta\leq\ 0.1^{o}$.
\end{resumen}
                  
\section{Introduction}
\label{S_intro}

The Universe is not transparent for gamma rays in the VHE range. Photon absorption in the intergalactic (IG) backgrounds is energy dependent \citep{1966PhRvL..16..252G} and starts to become substantial at GeV energies when cosmological distances are considered. In particular, VHE gamma rays coming from AGN can interact with photons in the IR-UV range present in the Extragalactic Background Light (EBL), producing electron-positron pairs. These pairs carry most of the energy from the original photons and can interact as well with IG photons via Inverse Compton scattering, boosting them to energies in the HE (high energy)-VHE range (30 MeV-100 TeV). This triggers a cascade process that converts the initial VHE photons into photons of lower energies which can travel further. Moreover, if the Intergalactic Magnetic Field (IGMF) is sufficiently strong, the bending effect on the electron-positron pair trajectories results in an extended isotropic emission of photons around the source. From Earth, one would see the source surrounded by an isotropically emitting {\it halo} \citep{1994ApJ...423L...5A}. A positive detection of this effect would have important implications on  the determination of currently uncertain cosmological parameters, such as the densities of the IG backgrounds and the intensity of the IGMF. 

Since the effect was proposed, several groups have tried to observe it using different methods \citep{2001A&A...366..746A, 2010A&A...524A..77A, 2010tsra.confE.192F}. In all cases only upper limits were calculated. In this work we study the possible extended emission in the blazar Markarian 421. For this we used VERITAS observations from the source in a quiescent state and compared them with observations from Markarian 421 April 2013 flare.
\section{Methods}

There are a few ways to search for extended emission and they all compare the measured angular distribution of excess events ($\theta^2$) to the one expected from a point source (PSF). We chose to exploit the fact that during a flare period the observations can be considered as coming from a ``point-like" source. This is because the pair-halo contribution from this period would take much longer to reach the observer than the direct photons from the source, resulting in a high flux ratio between direct and extended emission. In this way, we used the flare data to construct a template of the halo-free $\theta^2$ distribution.
The method for comparing both sets of data (quiescent and flare) goes as follows: in the first step we subtract the background and get the excess $\theta^2$ distributions. If we forget about the halo component for a moment, then we are left with the PSF shape for each corresponding set of data. If the PSF is the same (this is a challenge itself given that the PSF depends on many things: system PSF, spectral slope, azimuth, mispointings, etc.), then one $\theta^2$ plot should be the scaled version of the other (with the scaling factor being the fraction of gamma rays of each distribution). Now, if a halo component is present, then the $\theta^2$ distributions will be different and, after scaling, the shape of the distributions will not match to each other. For comparing the samples we used the $\chi^2$ test as follows: 

\begin{equation}
\chi^2=\sum_{i}\frac{\lbrace(N^i_{NOflare}-B^i_{NOflare})- F*(N^i_{flare}-B^i_{flare}))\rbrace^2}{{N^i_{NOflare}+B^i_{NOflare}+ F^2(N^i_{flare}+B^i_{flare})}}
\label{eq:chi}
\end{equation}

where $N^i_{NOflare}$ and $N^i_{flare}$ are the contents of the i-bin of each ON source distribution, and $B^i_{NOflare}$ and $B^i_{flare}$ are the corresponding estimations of the background distributions. $F$ is the scale factor applied to the Flare distribution, calculated from the integral of the distributions between the same limits considered in the calculation of $\chi^2$. To clarify, in equation \ref{eq:chi} the numerator is the ``halo excess" and the denominator its variance. We actually  end up using the reduced $\chi^2$ values: $\chi^2_{\nu}=\chi^2/(\nu-1)$, where $\nu$ is the total number of bins considered in the calculation ({\it d.o.f}).

\section{Results \& Discussion}

\begin{figure}[here]	
\centering
	\includegraphics[scale=0.265]{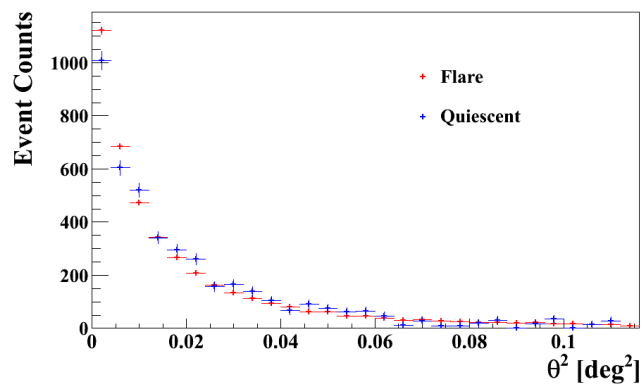}
	\caption{\footnotesize Scaled excess $\theta^2$ distributions for last April flare (red) and for the quiescent state of Mrk 421 (blue) during the same season.}
	\label{fig:Super}
	\end{figure}

We used 9.6 hours of high quality observations from Markarian 421 in a quiescent state taken from January to April 2013, a few days before the source entered a high state period, or flare. We also used 8.6 hours of high quality observations from the flare. In Figure \ref{fig:Super} we show the excess $\theta^2$ distributions obtained after scaling and up to around $\theta^2$=0.1, which is a reasonable range where the halo effect is expected to be present.

\begin{figure}[h]
	\centering
	\includegraphics[scale=0.265]{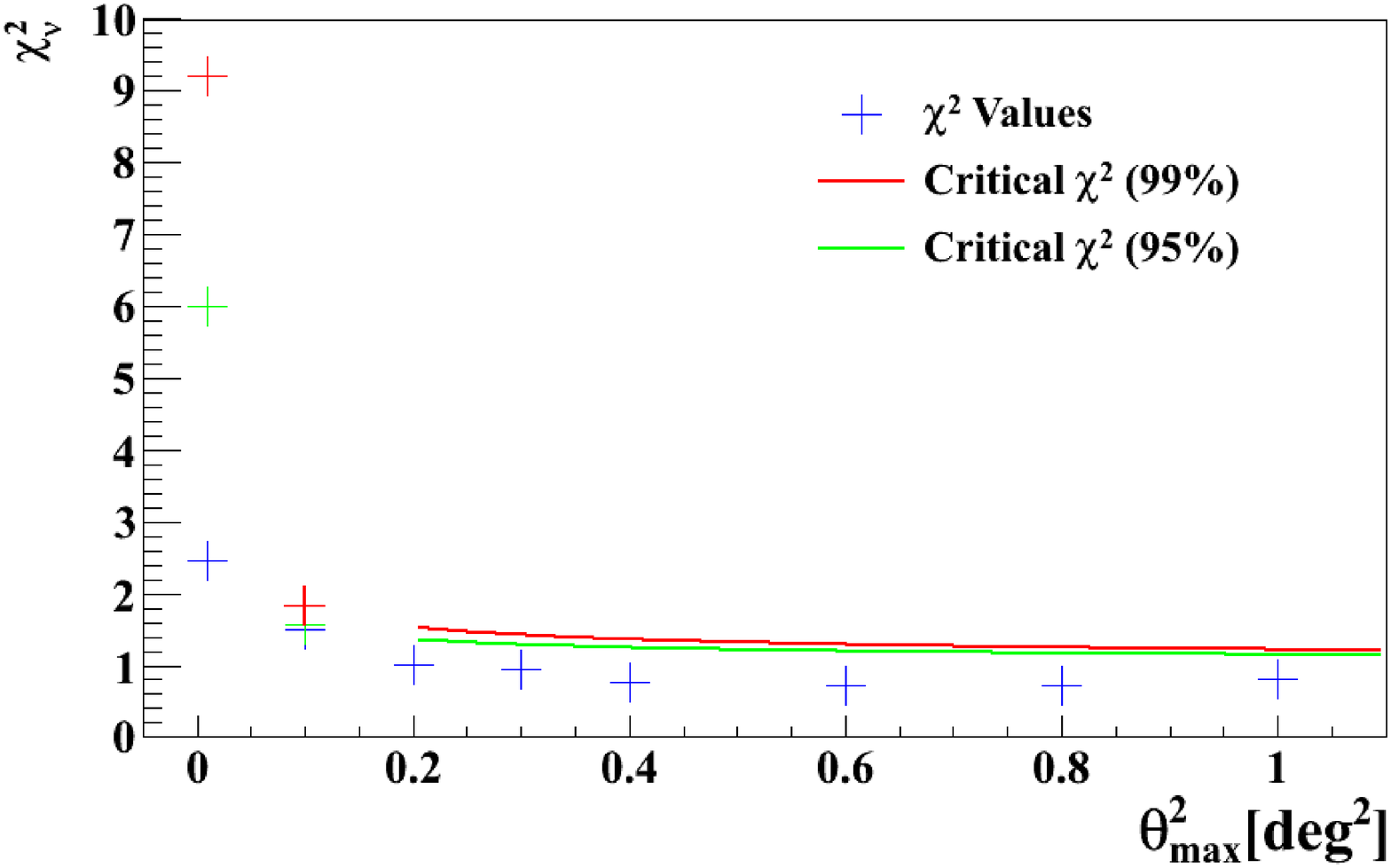}
	\caption{\footnotesize $\chi^2_{\nu}$ values (blue crosses) calculated for different angle ranges, from $\theta^2$=0$^{o}$ to $\theta^2_{max}$. The green (red) curve shows the critical $\chi^2$ values corresponding to 95\% (99 \%) CL. These curves cease to have validity for small number of bins ($\nu<30$), so for $\theta^2_{max}$=0.1$^{o}$ and $\theta^2_{max}$=0.01$^{o}$ we plot the corresponding $\chi^2_{\nu}$ values found in tables.}
	\label{fig:chi}
	\includegraphics[scale=0.265]{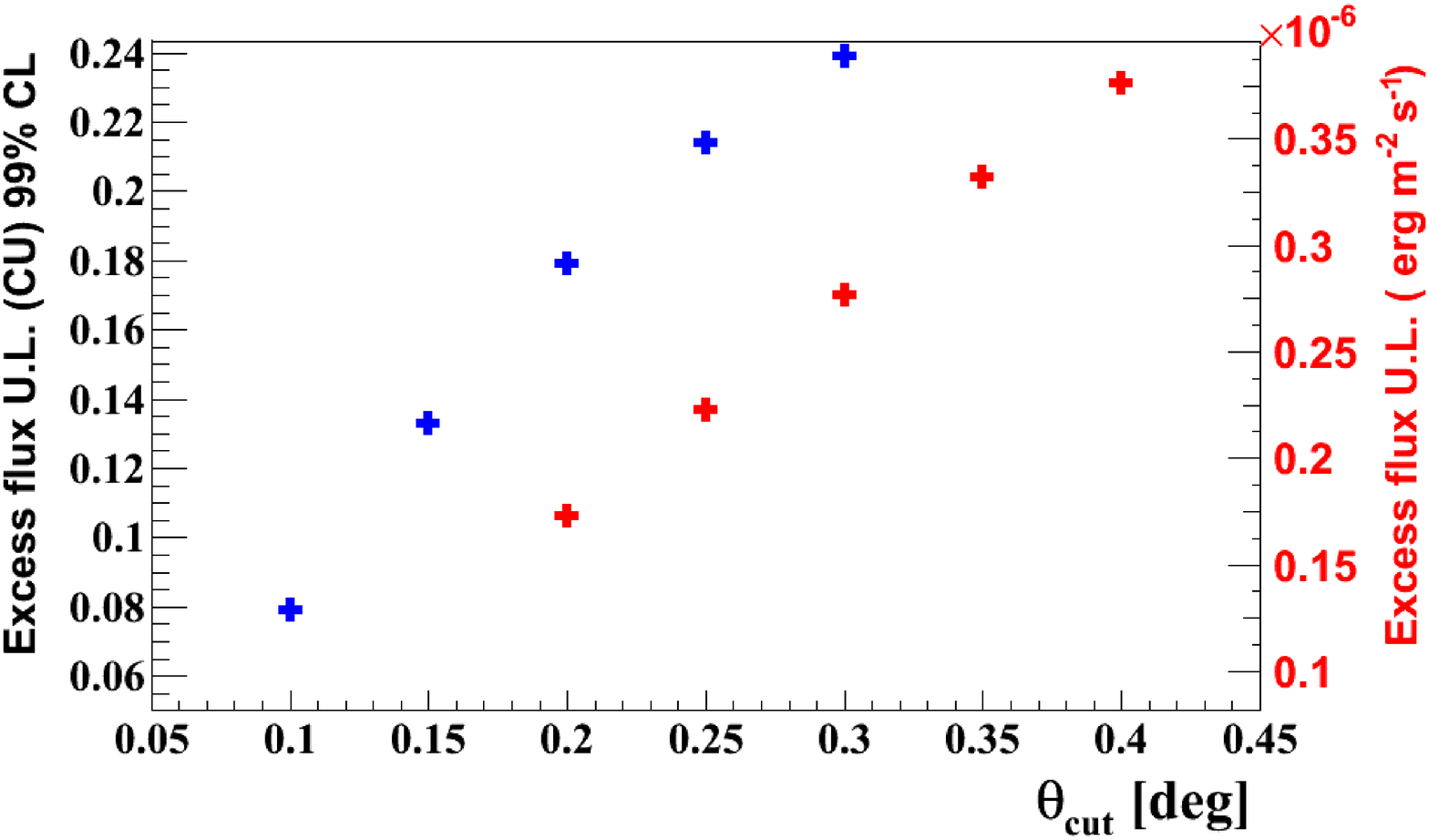}
	\caption{\footnotesize Upper limits on the flux of the extended emission in C.U and erg.m$^{-2}$s$^{-1}$ for different angle ranges: $0^{o}\leq\theta\leq\theta_{cut}$ (blue crosses) and $0.1^{o}\leq\theta\leq\theta_{cut}$ (red crosses).}
	\label{fig:Flux}
	\end{figure}

To apply the $\chi^2$ test we considered different ranges from $\theta^2=0$ to $\theta^2=\theta^2_{max}$. In Figure \ref{fig:chi} we show the calculated reduced $\chi^2$ values ($\chi^2_\nu$) as a function of $\theta^2_{max}$ together with the critical values corresponding to 99\% and 95\% confidence levels (CL). $\chi^2_\nu$ values below the critical curves can be considered as coming from distributions with no significant statistical difference at the corresponding CL. We see here that there is correlation between both samples for the considered $\theta^2$ ranges.

We conclude that there is no detectable halo effect in the sample for the ranges of $\theta^2$ considered. Therefore, we proceed to calculate upper limits for the halo flux and for different angle ranges. The calculations were done following Helene's method \citep{1983NIMPR.212..319H} with a CL of 99\% which corresponds to the detection threshold of 5$\sigma$. Results are shown in Figure \ref{fig:Flux}.
The most stringent upper limit is of $<$ 8\% C.U (1 C.U = 1.30$\times 10^{-6} \pm$ 1.99$\times 10^{-8}$m$^{-2}$s$^{-1}$ ), in the range $0^{o}\leq\theta\leq\ 0.1^{o}$ and $<$ 11\% C.U in the range $0.1^{o}\leq\theta\leq0.2^{o}$. This results are comparable to the ones reported in Aleksic et al. (2010), where they calculate upper limits for the same source using a different method and CL.

\acknowledgements
{We thank Dr. D.A. Supanitsky for useful discussions. This research is supported by grants from the U.S. Department of Energy Office of Science, the U.S. National Science Foundation and the Smithsonian Institution, by NSERC in Canada, by Science Foundation Ireland (SFI 10/RFP/AST2748) and by STFC in the U.K. We acknowledge the excellent work of the technical support staff at the Fred Lawrence Whipple Observatory and at the collaborating institutions in the construction and operation of the instrument.}

\bibliographystyle{baaa}
\bibliography{biblio}

\begin{thebibliography}{}

\bibitem[\protect\citeauthoryear{{Aharonian}, {Akhperjanian}, {Barrio},
  {Bernl{\'o}hr}, {Bolz} \& {B{\'o}rst}}{{Aharonian}
  et~al.}{2001}]{2001A&A...366..746A}
{Aharonian} F.~A.,    et~al., 2001, \aap, 366, 746

\bibitem[\protect\citeauthoryear{{Aharonian}, {Coppi} \& {Voelk}}{{Aharonian}
  et~al.}{1994}]{1994ApJ...423L...5A}
{Aharonian} F.~A.,  {Coppi} P.~S.,    {Voelk} H.~J.,  1994, \apjl, 423, L5

\bibitem[\protect\citeauthoryear{{Aleksi{\'c}}, {Antonelli}, {Antoranz},
  {Backes}, {Baixeras} \& {Barrio}}{{Aleksi{\'c}}
  et~al.}{2010}]{2010A&A...524A..77A}
{Aleksi{\'c}} J.,    et~al., 2010, \aap, 524, A77

\bibitem[\protect\citeauthoryear{{Fallon}, {Antonelli}, {Antoranz}, {Backes},
  {Baixeras} \& {Barrio}}{{Fallon} et~al.}{2010}]{2010tsra.confE.192F}
{Fallon} L.,    et~al., 2010, in 25th Texas Symposium on Relativistic
  Astrophysics. {}

\bibitem[\protect\citeauthoryear{{Gould} \& {Schr{\'e}der}}{{Gould} \&
  {Schr{\'e}der}}{1966}]{1966PhRvL..16..252G}
{Gould} R.~J.,  {Schr{\'e}der} G.,  1966, Physical Review Letters, 16, 252

\bibitem[\protect\citeauthoryear{{Helene}}{{Helene}}{1983}]{1983NIMPR.212..319%
H}
{Helene} O.,  1983, Nuclear Instruments and Methods in Physics Research, 212,
  319

\end{thebibliography}

\end{document}